\documentclass[acmsmall,screen]{acmart}
\renewcommand\footnotetextcopyrightpermission[1]{} 
\fancypagestyle{standardplain}{%
  \fancyhf{}%
  \fancyfoot[C]{\thepage}%
}
\fancypagestyle{plain}{%
  \fancyhf{}%
  \fancyfoot[C]{\thepage}%
}
\usepackage{tikz}
\usetikzlibrary{shapes, positioning, shapes.callouts, calc}
\usepackage{fontawesome5} 
\usepackage{multirow}
\usepackage{caption}
\usepackage{listings}
\usepackage{float}
\newfloat{listing}{tbhp}{lol}
\floatname{listing}{Listing}
\usepackage{xcolor}
\definecolor{MyGreen}{HTML}{81B365}
\definecolor{MyPurple}{HTML}{9773A6}
\usepackage{colortbl}
\usepackage{algorithm}
\usepackage{algpseudocode}
\usepackage{enumitem}
\usepackage{amsmath}
\usepackage{listings}
\usepackage{threeparttable}
\usepackage{makecell}
\usepackage{xcolor} 
\usepackage[most]{tcolorbox}
\definecolor{background}{HTML}{FFFFF5}
\definecolor{edge}{HTML}{87C2B1}
\newtcolorbox{mybox}{colback=background!35,
    colframe=MyGreen!80, 
    width=\columnwidth,
    arc=1mm,
    auto
    outer
    arc
}
\usepackage{placeins}
\AtBeginDocument{%
  }

\setcopyright{none}

\begin{document}

\fancypagestyle{standardplain}{%
  \fancyhf{}%
  \fancyfoot[C]{\thepage}%
}
\fancypagestyle{plain}{%
  \fancyhf{}%
  \fancyfoot[C]{\thepage}%
}
\pagestyle{plain}

\title{InspectCoder: Dynamic Analysis-Enabled Self Repair through Interactive LLM-Debugger Collaboration}

\author{Yunkun Wang}

\affiliation{%
  \institution{Zhejiang University}
  \city{Hangzhou}
  \country{China}}
\email{wangykun@zju.edu.cn}

\author{Yue Zhang}
\affiliation{%
  \institution{Alibaba Group}
  \city{Hnagzhou}
  \country{China}
}
\email{shiyu.zy@alibaba-inc.com}

\author{Guochang Li}
\affiliation{%
  \institution{Zhejiang University}
  \city{Hangzhou}
  \country{China}}
\email{gcli@zju.edu.cn}

\author{Chen Zhi}
\affiliation{%
 \institution{Zhejiang University}
 \city{Hangzhou}
 \country{China}}
\email{zjuzhichen@zju.edu.cn}

\author{Binhua Li}
\affiliation{%
  \institution{Alibaba Group}
  \city{Hangzhou}
  \country{China}}
\email{binhua.lbh@alibaba-inc.com}

\author{Fei Huang}
\affiliation{%
  \institution{Alibaba Group}
  \city{Hangzhou}
  \country{China}}
\email{f.huang@alibaba-inc.com}

\author{Yongbin Li$^{\dagger}$}
\affiliation{%
  \institution{Alibaba Group}
  \city{Hangzhou}
  \country{China}}
\email{shuide.lyb@alibaba-inc.com}

\author{Shuiguang Deng$^{\dagger}$}
\affiliation{%
  \institution{Zhejiang University}
  \city{Hangzhou}
  \country{China}}
\email{dengsg@zju.edu.cn}

\renewcommand{\thefootnote}{}
\footnotetext{$^{\dagger}$Corresponding authors.}
\renewcommand{\thefootnote}{\arabic{footnote}}

\begin{abstract}
Large Language Models (LLMs) frequently generate buggy code with complex logic errors that are challenging to diagnose. While existing LLM-based self-repair approaches conduct intensive static semantic analysis or reply on superficial execution logs, they miss the in-depth runtime behaviors that often expose bug root causes—lacking the interactive dynamic analysis capabilities that make human debugging effective.

We present InspectCoder, the first agentic program repair system that empowers LLMs to actively conduct dynamic analysis via interactive debugger control. Our dual-agent framework enables strategic breakpoint placement, targeted state inspection, and incremental runtime experimentation within stateful debugger sessions. Unlike existing methods that follow fixed log collection procedures, InspectCoder adaptively inspects and perturbs relevant intermediate states at runtime, and leverages immediate process rewards from debugger feedback to guide multi-step reasoning, transforming LLM debugging paradigm from blind trial-and-error into systematic root cause diagnosis.

We conduct comprehensive experiments on two challenging self-repair benchmarks: BigCodeBench-R and LiveCodeBench-R. InspectCoder achieves 5.10\%–60.37\% relative improvements in repair accuracy over the strongest baseline, while delivering 1.67x-2.24x superior bug-fix efficiency respectively. We also contribute InspectWare, an open-source middleware that abstracts debugger complexities and maintains stateful debugging sessions across mainstream Python testing frameworks. Our work provides actionable insight into the interactive LLM-debugger systems, demonstrating the significant potential of LLM-driven dynamic analysis for automated software engineering.
\end{abstract}



\maketitle

\thispagestyle{plain}
\pagestyle{plain}

\section{Introduction}
\begin{figure}[thbp]
    \centering
    \includegraphics[width=1\linewidth]{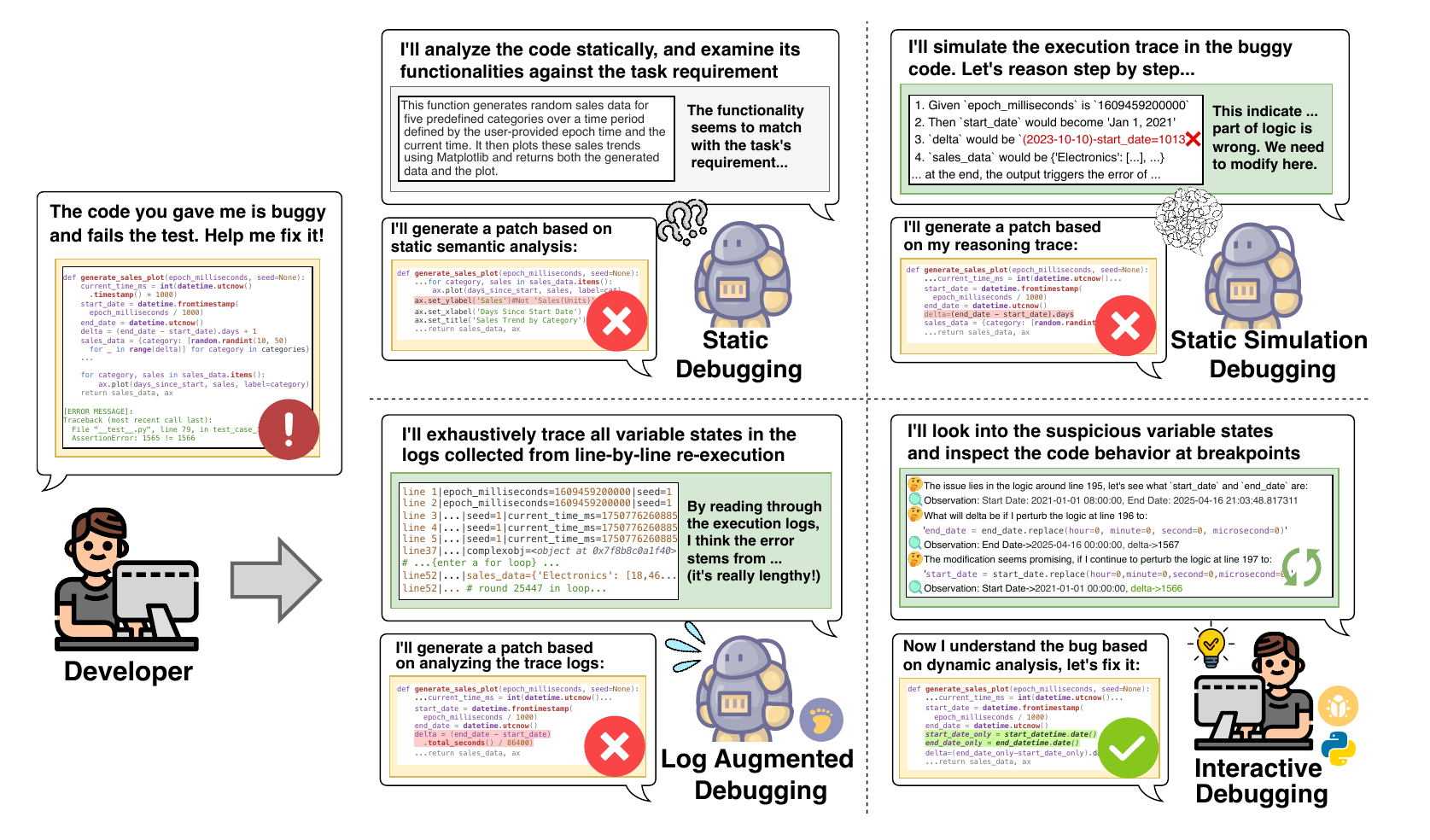}
    \caption{Comparison of different LLM-based program repair paradigms. Green boxes denote the extent of dynamic semantic understanding accessible to LLMs.}
    \label{fig:inspectcoder_motivated_example}
\end{figure}

Large Language Models (LLMs) have demonstrated remarkable code generation capabilities~\cite{LLM-codegeneration,codescope,swebench}, yet they frequently produce buggy code with hidden discrepancies between requirements and implementations~\cite{LLM-generated-bugs,25wehre-LLM-bug}. To address this issue, recent research has explored the self-repair capabilities of LLMs ~\cite{23Self-Edit,23self-refine,24Self-Repair,24Self-Debug,25KnowledgeAPR}, enabling them to iteratively debug and refine their self-generated code. Current program repair approaches primarily rely on LLMs' pre-trained understanding of static program semantics. However, dynamic program analysis—such as interactive program inspection with debuggers—remains a neglected aspect in the field.

As illustrated in Figure~\ref{fig:inspectcoder_motivated_example}, given the buggy code and error test messages, the majority of existing LLM-based program repair approaches adopt a \textbf{Static Debugging} paradigm \cite{23Self-Edit, 24Self-Repair, 23ChatRepair, 25RepairAgent}, where LLMs comprehend the error message primarily through syntactic and static semantic analysis, while overlooking the intermediate runtime behaviors.
A subset of these approaches \cite{24Self-Debug, 24FixAgent, PATCH} employ the \emph{rubber duck debugging}~\cite{rubberduck}, which mandates line-by-line code explanation prior to patch generation, thereby capturing fine-grained static semantics gaps.
To move beyond the superficial understanding, some recent efforts \cite{24Self-Debug, 24NExT}, which we categorize as \textbf{Static Simulation Debugging}, prompt LLMs to verbally simulate the program's execution trace. These techniques leverage such ``pseudo-dynamic'' information to improve patch quality without accessing real runtime data.
In contrast, more recent methods \cite{24LDB, 25AutoSD} have introduced external procedures to automatically collect dynamic information for LLMs. 
We refer to them as \textbf{Log Augmented Debugging}. For instance, LDB \cite{24LDB} performs indiscriminate instrumentation to record all variable values and gathers execution logs. AutoSD \cite{25AutoSD} iteratively make full re-execution with templated logging scripts and prompts LLM to produce explainable output. However, they do not provide any support for LLM-initiated interactions that enable flexible inspection and perturbations on programs' runtime dynamics.

We identify two fundamental limitations in these repair approaches:
\textbf{(1) Insufficient understanding of runtime behavior.}
As shown in Figure \ref{fig:inspectcoder_motivated_example}, \emph{Static methods} inherently have limited access to dynamic information, tending to identify obvious mistakes while missing complex logic errors~\cite{24leap2patch}.
Despite \emph{Static Simulation Debugging} tries to reason over execution behavior, recent studies \cite{24CruxEval,25CodeReason,24CodeMind} suggest that LLMs possess limited dynamic understandings in code, and hallucinated reasoning trace may oppositely degrade patch quality. 
While \textit{Log Augmented Debugging} introduces actual runtime data, the collected information is often superficial, overly verbose, and lacks semantic relevance, resulting in lengthy trace logs that can overwhelm the LLMs. Furthermore, these approaches struggle with in-depth runtime states (eg. nested data structures with internal attributes or compound expressions) that require context-aware inspection. In contrast, human developers utilize debuggers to selectively examine pertinent variables or expressions, enabling focused and accurate understanding of program behavior.
\textbf{(2) Absence of incremental debugging validation.} 

Existing program repair methods rely solely on final test outcomes (and log outputs) to assess the correctness of generated patches, requiring the model to complete an entire repair attempt before receiving any validation feedback. This irreversible code changes with \textit{outcome-only reward} risks guiding the model toward fundamentally flawed repair paths \cite{24processreward,25astray-swe-prm}. In contrast, human developers typically adopt an incremental debugging strategy \cite{25human-incremental-debug, 17debug-practitioner}: using debugger tools to interactively perturb critical runtime states within a temporal session\footnote{For example, interactive state manipulation in PDB: \url{https://docs.python.org/3/library/pdb.html\#pdbcommand-interact}}, and immediately observe the resulting behavioral changes. This provides a real-time \textit{process reward}~\cite{25astray-swe-prm} that enables multi-step reasoning at one suspicious program state and early adjustment of bug understandings, while avoiding irreversible code changes and costly full re-execution.

Inspired by human debugging practice, we propose InspectCoder, an agentic system that diagnoses and fixes LLM-generated bugs by actively inspecting their runtime behaviors with interactive debugger tools. 
Specifically, InspectCoder adopts a dual-agent framework, consisting of (1) a Program Inspector agent that interacts with debugger tools to conduct in-depth dynamic analysis and identify bug root causes and
(2) a Patch Coder agent that produces patches based on identified root causes and validates patches by executing test cases. Emulating human debugging workflows, InspectCoder enables flexible breakpoint placement, targeted state inspection, and incremental behavior exploration under runtime code modifications. To enable robust and efficient interaction between LLMs and state-transitioning debugger environments, we further develop InspectWare, 
a middleware that encapsulates complex debugger protocols and maintains a stateful debugging context to support multi-turn LLM reasoning and action planning.

We conduct comprehensive evaluations on two challenging self-repair benchmarks, BigCodeBench-R and LiveCodeBench-R, which contain LLM-generated bugs on high-quality programming tasks featuring detailed code requirements in diverse programming scenarios. We conduct experiments with SOTA LLMs with minimum data contaminent risks, including Qwen2.5-Max~\cite{qwen25}, DeepSeek-V3~\cite{deepseek-v3}, GPT-4o~\cite{gpt4o}, and Claude-3.5~\cite{claude35}. Experimental results demonstrate that InspectCoder exhibits superior fixing capacities, achieving 5.10\% and 60.37\% relative improvements in resolve rate over the strongest established baseline, while exhibiting 1.67x and 2.24x superior bug-fix efficiency respectively. We also investigate various InspectCoder's design choices and behavioral patterns of LLM-driven dynamic analysis, offering practical insights for incorporating interactive debugging into LLM-based program repair systems.

In summary, the main contributions of this work are as follows:

\begin{itemize}[leftmargin=0.4cm] 
\item \textbf{A Novel Agent-Based Program Repair Framework:} We present InspectCoder, the first agentic program repair system that empowers LLMs to conduct dynamic analysis through flexible debugger interaction. InspectCoder gives an LLM agent the initiative to strategically place breakpoints, inspect runtime state, make incremental runtime experiments, and leverage immediate debugger feedback as process rewards to guide root cause identification. This marks a paradigm shift from passive consumption of post-hoc logs to active LLM-driven runtime analysis in program repair.

\item \textbf{Comprehensive Evaluation and Insights:} We conduct extensive evaluations across four SOTA LLMs on two challenging benchmarks, BigCodeBench-R and LiveCodeBench-R. InspectCoder achieves 5.10\%-60.37\% relative resolve rate increases and 1.67x-2.24x efficiency gains over the strongest baseline. Our empirical studies provide actionable insights into design choices for stateful LLM-debugger interactions and patterns of LLM-driven dynamic analysis system.

\item \textbf{An Open LLM-Debugger Middleware:} We develop and release InspectWare, a standardized middleware that abstracts debugger complexities and maintains stateful debugging sessions for LLMs. InspectWare supports comprehensive debugger operations across mainstream Python testing frameworks (unittest, pytest, competition-style IO tests), enabling broader adoption of dynamic analysis in practical LLM-based program repair workflows. Our code will be available at \url{https://github.com/greenlight2000/InspectCoder_framework}.
\end{itemize}

\section{Approach}
\begin{figure*}[t]
    \centering
    \includegraphics[width=1.0\linewidth]{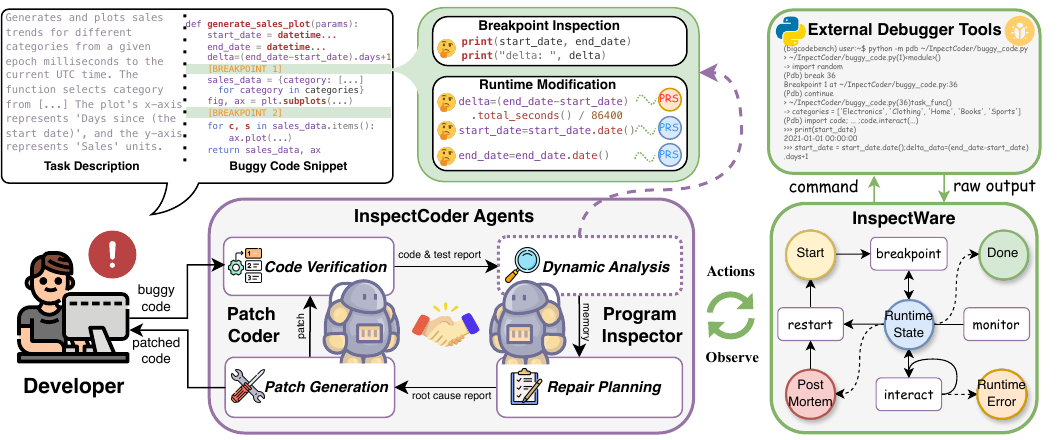}
    \caption{An Overview of InspectCoder framework. InspectCoder Agents interact with external debugger tool through InspectWare. Each Inspector action modifies the debugger session state for subsequent operations.}
    \label{fig:inspectcoder_framework}
\end{figure*}

\subsection{Task Definition}
\label{sec:task_definition}

We formulate the task of LLM-based program repair as follows. Given a code requirement $R$, failing test case $t_{\text{fail}}$, and a buggy implementation $C_{\text{buggy}}$ initially synthesized by an LLM $\theta$ from $R$, the objective is to produce a corrected program $C_{\text{patched}}$ that passes all test cases $T$. For benchmarks with public-private test splits, we use only public tests during repair and full tests for validation.

\begin{equation}
\label{eq:self_repair}
\begin{aligned}
C_{\text{patched}} &\leftarrow \text{Debug}_{\theta}(R, C_{\text{buggy}}, t_{\text{fail}}) \\
\text{s.t.} \quad\quad \forall t_i &\in T,~ f_{\text{test}}(C_{\text{patched}}, t_i) = 1, \\
    \text{where } C_{\text{buggy}} &\sim \text{Synthesis}_\theta(R), \ t_{\text{fail}} \in T.
\end{aligned}
\end{equation}

\noindent In alignment with recent self-repair work~\cite{24LDB, 25KnowledgeAPR, 25AutoSD}, we target LLM-generated bugs in challenging function synthesis. It is important to note that self-repair addresses a complementary challenge to repository issue-solving tasks like SWE-bench~\cite{swebench}, which emphasizes exploring repositories intents and inferring underspecified requirements from bug reports. We instead focus on diagnosing hidden semantic gaps where LLM implementations deviate from explicit requirements, which challenges LLMs to correct their own cognitive biases. This setting restores applicable scenarios for dynamic analysis, and also mirrors practical workflows where developers handle LLM-generated functional errors~\cite{22human-debug-function-error}. Section 6 further discusses how InspectCoder can be integrated in repository issue-solving workflows.

\subsection{Overview of InspectCoder framework}
We present InspectCoder, a novel dual-agent system that diagnoses and fixes buggy programs by actively inspecting the program runtime behavior.  As illustrated in Figure \ref{fig:inspectcoder_framework}, InspectCoder comprises two specialized agents that work in tandem within an iterative debugging loop. The \textbf{Program Inspector} employs reactive dynamic analysis to systematically control program execution and runtime states through a LLM-debugger middleware named \textbf{InspectWare}, while the \textbf{Patch Coder} leverages the Inspector's insights to generate and verify code patches. 
This separation of tasks allows each agent to focus on its expertise while maintaining effective communication through helpful information exchange.

\begin{center}
\begin{minipage}{0.95\columnwidth}
\begin{lstlisting}[
 basicstyle=\footnotesize\ttfamily,
 breaklines=true,
 breakindent=8pt,
 showstringspaces=false,
 rulecolor=\color{MyPurple},
 framerule=1pt,
 escapeinside={(*}{*)},
 caption={Prompt design in Program Inspector.},
 label={fig:inspector_prompt}
]
<<--System_Prompt-->>
You are a senior developer helping to debug a buggy code. You will be given the intended Coding Task, the Buggy Code, and Test Results of buggy code.
Your goal is to inspect and analyze the current buggy code to gain insights so that you can **propose an repair plan** to repair code.
PDB terminal serves to pause the program exeution at certain code lines for inspection. You can interatively interact with a PDB terminal to debug. Here are some tools you can use:
{Tool_explanations}
At each iteration, you should first **think** step by step then take **action** by calling above tools in a ```...``` block, and I will provide the **observation** of your action. Please respond with one "### THOUGHT" and one "### ACTION", here are some typical examples: {Few-shots} 
{More_Restrictions}

<<--User_Prompt-->>
## Coding Task\n{coding_task}                          // (*$R$*)
## Buggy Code\n{buggy_code}                            // (*$C_\text{buggy}$*)
## Test Results
Error Message on Failing Test Case:\n{test_case}       // (*$t_\text{fail}$*)
## PDB Session State
{InspectWare_Monitored_Session_States}                 // (*$\leftarrow$*) InspectWare
----
{agentic_reasoning_trajectory}       // Accumulated thought-action-observation

\end{lstlisting}
\end{minipage}
\end{center}

\subsection{Program Inspector: Interactive Debugging for Dynamic Analysis}
The Program Inspector serves as the analytical core of our system, employing a ReAct framework \cite{23react} to autonomously inspect program state. Unlike traditional static debugging or log-augmneted debugging methods, our Inspector can interact with a live debugger environment through our `InspectWare' middleware, actively gathering dynamic information to diagnose the bug's root cause. 
We demonstrate the overall prompt of Program Inspector in Listing~\ref{fig:inspector_prompt}.

\subsubsection{Action Space}

To facilitate flexible LLM-debugger communication, we designed a primitive yet powerful action space that abstracts the fundamental operations of human debugger usage. Each Inspector initiated action will be translated into debugger commands in InspectWare and returned as meaningful feedback observations.

\begin{itemize}[leftmargin=0.4cm]
\item \textbf{\texttt{set\_breakpoint(line: int)}}: Places a breakpoint at a specific line number in current file, allowing the Inspector to pause execution at a point of interest before the line is executed. A breakpoint will be removed by InspectWare automatically once being triggered, which relieves the Inspector agent from the overhead of managing breakpoint states.
\item \textbf{\texttt{control\_execution(cmd: str)}}: Provides navigational control over the debugging process. The supported commands are (1) \textbf{\texttt{continue}}, which resumes execution until the next breakpoint is triggered, and (2) \textbf{\texttt{restart}}, which resets the debugging session to the first line. This is useful when the Inspector needs to go back to an earlier line in the code, or recover from post mortem errors during interaction.
\item \textbf{\texttt{interact\_code(code: str)}}: This is the Inspector's primary action for dynamic analysis and interactions. InspectWare will open an interactive Python interpreter within the paused execution context, allowing the agent to execute arbitrary code and gain immediate outputs. The Inspector can print variables to observe current states or inject perturbation logic to validate potential program behaviors. Crucially, exceptions raised within this interactive session do not crush the execution, enabling continuous reasoning at anchor points and efficient runtime debugging.
\item \textbf{\texttt{propose\_repair(plan: str)}}: This action summarizes the accumulated insights and terminates the dynamic analysis process. Program Inspector can actively invokes this action to share the identified root cause with Patch Coder to propose a repair.
\end{itemize}
To accelerate the reasoning process and ensure logical consistency among action invocations, we allow the Inspector to generate thoughts and a sequence of actions in a single turn. All actions are formatted as API calls, e.g. \texttt{action\_name(parameter)}, for reliable parsing and execution.

\subsubsection{Dynamic Analysis Strategies}
While debugger integration is intuitive for enabling dynamic analysis, it faces a critical challenge: \textbf{debuggers are complex tools requiring strategic orchestration of primitive actions}, yet LLMs lack training data on agentic debugger usage and struggle to compose these actions effectively without guidance. To address this cold-start problem, we derive two complementary debugging strategies from existing empirical studies on human debugging behaviors~\cite{testhypothesis,forwardreason} and encode them as few-shot instruction learning. This enables LLMs to autonomously select and compose appropriate action sequences for effective dynamic analysis.

\begin{figure}[htbp]
\centering
\begin{tikzpicture}
    \node[
        rectangle,
        text width=0.34\columnwidth,
        inner sep=8pt,
        anchor=north west,
        opacity=0
    ] (thought_measure) at (0,0) {
        \textcolor{blue!70}{\small\faLightbulb~\textbf{THOUGHT}}\\[4pt]
        \footnotesize\textit{From previous observation and analysis, I suspect the bug is related to [variable]. [expression] should be [value], I need to inspect related variable's value to examine if discrepancy exists.}
    };
    
    \node[
        rectangle,
        text width=0.37\columnwidth,
        inner sep=8pt,
        anchor=north east,
        opacity=0
    ] (action_measure) at (0.82\columnwidth,0) {
        \textcolor{gray!70}{\small\faCode~\textbf{ACTION}}\\[4pt]
        \footnotesize\ttfamily
        \verb|```|toolcalls\\
        set\_breakpoint(11)\\
        control\_execution("continue")\\
        interact\_code("print(obj\_b.c['key'])")\\
        interact\_code("print(var\_a)")\\
        \verb|```|
    };
    
    \path let \p1 = (thought_measure.north), \p2 = (thought_measure.south),
              \p3 = (action_measure.north), \p4 = (action_measure.south) in
        \pgfextra{
            \pgfmathsetmacro{\heightone}{\y1-\y2}
            \pgfmathsetmacro{\heighttwo}{\y3-\y4}
            \pgfmathsetmacro{\maxheight}{max(\heightone,\heighttwo)}
            \xdef\myheight{\maxheight pt}
        };
    
    \node[
        rectangle,
        rounded corners=3pt,
        fill=blue!5,
        draw=blue!40,
        line width=0.8pt,
        text width=0.34\columnwidth,
        inner sep=8pt,
        anchor=north west,
        minimum height=\myheight,
        align=left
    ] (thought) at (0,0) {
        \textcolor{blue!70}{\small\faLightbulb~\textbf{THOUGHT}}\\[4pt]
        \footnotesize\textit{The error suggests obj\_b.c['key'] has an unexpected value. I will inspect it at line 11, then trace upstream to var\_a to understand how this erroneous value originated.}
    };
    
    \node[
        rectangle,
        rounded corners=3pt,
        fill=gray!3,
        draw=gray!50,
        line width=0.8pt,
        text width=0.37\columnwidth,
        inner sep=8pt,
        anchor=north east,
        minimum height=\myheight,
        align=left
    ] (action) at (0.82\columnwidth,0) {
        \textcolor{gray!70}{\small\faCode~\textbf{ACTION}}\\[4pt]
        \footnotesize\ttfamily
        \verb|```|debugger\\
        set\_breakpoint(11)\\
        control\_execution("continue")\\
        interact\_code("print(obj\_b.c['key'])")\\
        interact\_code("print(var\_a)")\\
        \verb|```|
    };
\end{tikzpicture}
\caption{Demonstration for Dynamic Debugging Strategy: Breakpoint Inspection.}
\label{fig:breakpoint_inspection}
\end{figure}
\paragraph{Breakpoint Inspection:} This strategy examines how erroneous runtime values originate from buggy lines and propagate to test oracles, building causal understanding of bug manifestation. By strategically placing breakpoints at suspicious locations, the Inspector collects rich dynamic information—such as variable values or arbitrary expressions for complex objects—to understand program behavior during runtime. As illustrated in Listing \ref{fig:breakpoint_inspection}, we guide the LLM through strategy demonstration to trace suspicious variables along with their upstream variables participating in value assignment. Through observing runtime states, the Inspector progressively identifies discrepancies between expected and actual program states at specific execution moments, ultimately enabling root cause identification~\cite{testhypothesis}.

\paragraph{Runtime Modification:} This strategy validates debugging hypotheses by experimentally perturbing runtime values or injecting potential patch logic at targeted breakpoints~\cite{forwardreason}. By initiating an interactive Python session within the paused execution context, the Inspector tests runtime code changes without modifying source files—a critical advantage over existing methods~\cite{24sweagent,25AutoSD} that require irreversible error-prone file editing and costly re-execution. Leveraging the debugger's interactive execution control, our approach enables temporal, reversible perturbations of intermediate program states, providing immediate process reward signals for early adjustment of root cause understanding. To our knowledge, this is the first work to incorporate such debugger capabilities for automated program repair.
We specifically demonstrate this strategy in two debugging scenarios:
\begin{figure}[htbp]
\centering
\begin{tikzpicture}
    \node[
        rectangle,
        text width=0.34\columnwidth,
        inner sep=8pt,
        anchor=north west,
        opacity=0
    ] (thought_measure) at (0,0) {
        \textcolor{blue!70}{\small\faLightbulb~\textbf{THOUGHT}}\\[4pt]
        \footnotesize\textit{I suspect necessary logic is missing before line 10. I can insert a debugging logic there and continue executing the rest code to observe how it affects program behavior.}
    };
    
    \node[
        rectangle,
        text width=0.37\columnwidth,
        inner sep=8pt,
        anchor=north east,
        opacity=0
    ] (action_measure) at (0.82\columnwidth,0) {
        \textcolor{gray!70}{\small\faCode~\textbf{ACTION}}\\[4pt]
        \footnotesize\ttfamily
        \verb|```|debugger\\
        set\_breakpoint(10); set\_breakpoint(40)\\
        control\_execution("continue")\\
        interact\_code("var\_a=perturb(var\_a)")\\
        control\_execution("continue")\\
        \verb|```|
    };
    
    \path let \p1 = (thought_measure.north), \p2 = (thought_measure.south),
              \p3 = (action_measure.north), \p4 = (action_measure.south) in
        \pgfextra{
            \pgfmathsetmacro{\heightone}{\y1-\y2}
            \pgfmathsetmacro{\heighttwo}{\y3-\y4}
            \pgfmathsetmacro{\maxheight}{max(\heightone,\heighttwo)}
            \xdef\myheight{\maxheight pt}
        };
    
    \node[
        rectangle,
        rounded corners=3pt,
        fill=blue!5,
        draw=blue!40,
        line width=0.8pt,
        text width=0.34\columnwidth,
        inner sep=8pt,
        anchor=north west,
        minimum height=\myheight,
        align=left
    ] (thought) at (0,0) {
        \textcolor{blue!70}{\small\faLightbulb~\textbf{THOUGHT}}\\[4pt]
        \footnotesize\textit{The error at line 40 suggests var\_a is unprocessed. I suspect missing logic before line 10. I will insert experimental logic there and continue execution to validate if this resolves the error.}
    };
    
    \node[
        rectangle,
        rounded corners=3pt,
        fill=gray!3,
        draw=gray!50,
        line width=0.8pt,
        text width=0.37\columnwidth,
        inner sep=8pt,
        anchor=north east,
        minimum height=\myheight,
        align=left
    ] (action) at (0.82\columnwidth,0) {
        \textcolor{gray!70}{\small\faCode~\textbf{ACTION}}\\[4pt]
        \footnotesize\ttfamily
        \verb|```|debugger\\
        set\_breakpoint(10);set\_breakpoint(40)\\
        control\_execution("continue")\\
        interact\_code("var\_a=perturb(var\_a)")\\
        control\_execution("continue")\\
        \verb|```|
    };
\end{tikzpicture}
\caption{Demonstration for Dynamic Debugging Strategy: Runtime Modification (insert missing logic).}
\label{fig:runtime_missing}
\end{figure}
\begin{enumerate}[label=\arabic*), leftmargin=0.4cm]
\item \emph{Scenario 1: Debugging on Missing Logic.} As shown in Listing \ref{fig:runtime_missing}, when the LLM suspects that missing logic causes errors in later execution, it can insert test logic at the suspected location and validate the effect by continuing execution to see whether the error still occurs.

\item \emph{Scenario 2: Debugging on Flawed Logic.} As shown in Listing \ref{fig:runtime_alternative}, when existing code logic is flawed, the modified program behavior cannot be observed through continued execution because the flawed code line cannot be bypassed. Therefore, we guide the LLM to validate its experimental replacement logic within the \texttt{interact\_code} session. Specifically, the LLM first navigates to the breakpoint where the logic is flawed, then executes replacement logic within the breakpoint context, along with subsequent logic to observe the behavior from live session feedback.
\end{enumerate}
\begin{figure}[htbp]
\centering
\begin{tikzpicture}
    \node[
        rectangle,
        text width=0.34\columnwidth,
        inner sep=8pt,
        anchor=north west,
        opacity=0
    ] (thought_measure) at (0,0) {
        \textcolor{blue!70}{\small\faLightbulb~\textbf{THOUGHT}}\\[4pt]
        \footnotesize\textit{I suspect existing logic at line 10 is flawed. I can start a python session before line 10 and validate how alternative logic affects subsequent code behavior inside the interactive session.}
    };
    
    \node[
        rectangle,
        text width=0.37\columnwidth,
        inner sep=8pt,
        anchor=north east,
        opacity=0
    ] (action_measure) at (0.82\columnwidth,0) {  
        \textcolor{gray!70}{\small\faCode~\textbf{ACTION}}\\[4pt]
        \footnotesize\ttfamily
        \verb|```|debugger\\
        set\_breakpoint(10)\\
        control\_execution("continue")\\
        interact\_code("alternative\_logic")\\
        interact\_code("subsequent\_logic")\\
        \verb|```|
    };
    
    \path let \p1 = (thought_measure.north), \p2 = (thought_measure.south),
              \p3 = (action_measure.north), \p4 = (action_measure.south) in
        \pgfextra{
            \pgfmathsetmacro{\heightone}{\y1-\y2}
            \pgfmathsetmacro{\heighttwo}{\y3-\y4}
            \pgfmathsetmacro{\maxheight}{max(\heightone,\heighttwo)}
            \xdef\myheight{\maxheight pt}
        };
    
    \node[
        rectangle,
        rounded corners=3pt,
        fill=blue!5,
        draw=blue!40,
        line width=0.8pt,
        text width=0.34\columnwidth,
        inner sep=8pt,
        anchor=north west,
        minimum height=\myheight,
        align=left
    ] (thought) at (0,0) {
        \textcolor{blue!70}{\small\faLightbulb~\textbf{THOUGHT}}\\[4pt]
        \footnotesize\textit{I suspect the logic at line 10 is flawed. Since I cannot bypass it, I will navigate to line 10, execute alternative logic in the interactive session, then run subsequent code to verify the fix.}
    };
    
    \node[
        rectangle,
        rounded corners=3pt,
        fill=gray!3,
        draw=gray!50,
        line width=0.8pt,
        text width=0.37\columnwidth,
        inner sep=8pt,
        anchor=north east,
        minimum height=\myheight,
        align=left
    ] (action) at (0.82\columnwidth,0) {  
        \textcolor{gray!70}{\small\faCode~\textbf{ACTION}}\\[4pt]
        \footnotesize\ttfamily
        \verb|```|debugger\\
        set\_breakpoint(10)\\
        control\_execution("continue")\\
        interact\_code("<alternative\_logic>")\\
        interact\_code("<validation\_logic>")\\
        \verb|```|
    };
\end{tikzpicture}
\caption{Demonstration: Runtime Modification (replace flawed logic).}
\label{fig:runtime_alternative}
\end{figure}

\subsubsection{Root Cause Report Generation}

Through in-depth dynamic analysis, the Program Inspector accumulates a comprehensive reasoning memory that captures dynamic insights from program behaviors. The Inspector will actively invokes the \texttt{propose\_repair} action when sufficient dynamic evidence has been gathered for root cause identification, or when the reasoning steps reach a preset limit. Upon invocation, the Inspector summarizes its reasoning memory into a root cause report that identifies where and why the bug occurs and proposes concrete repair plan. We prompt the LLM to generate following a structured patterns of "the root cause of the bug is that ..., to fix the bug, consider ...". This approach ensures the dynamic analysis trace can be condensed into meaningful and constructive guidance for patch generation.

\subsection{Patch Coder: Root Cause-Grounded Patch Synthesis}
The Patch Coder utilizes the root cause analysis insights to concrete code modifications. This component completes the InspectCoder framework by bridging dynamic analysis with program repair. 
We demonstrate the overall prompt of Patch Coder in Listing~\ref{fig:coder_prompt}.
\begin{center}
\begin{minipage}{0.95\columnwidth}
\begin{lstlisting}[
 basicstyle=\footnotesize\ttfamily,
 breaklines=true,
 breakindent=8pt,
 showstringspaces=false,
 rulecolor=\color{MyPurple},
 framerule=1pt,
 escapeinside={(*}{*)},
 caption={Prompt design in Patch Coder.},
 label={fig:coder_prompt}
]
<<--System_Prompt-->>
You are a expert python developer who is tasked to repair a buggy code for a coding task. You will be given the intended Coding Task, the Buggy Code, Error Message, an Edit Plan for the buggy code suggested by a debugger agent, history of your previous repair attempts (if any). Your goal is to generate a repaired code within ```...``` block.

<<--User_Prompt-->>
# Coding Task{coding_task}                             // (*$R$*)
## Code Version {version}\n```{buggy_code}\n```        // (*$C_\text{buggy}$*)
### Stack Trace and Error:\n{stack_trace_and_error}    // (*$t_\text{fail}$*)
### Repair Plan:\n{repair_plan}                        // (*$\leftarrow$*) Program Inspector

\end{lstlisting}
\end{minipage}
\end{center}
\subsubsection{Utilizing Dynamic Analysis Insights for Patching}  
Unlike many LLM-based program repair approaches that often leap to patch generation in the absence of sufficient information \cite{24leap2patch}, our Patch Coder leverages insightful root cause grounding to synthesize high-quality patches. 
Patch Coder operates by first receiving concise root cause reports from the Program Inspector, which distill key insights from dynamic analysis. Using these reports along with the task descriptions and buggy code, it generates dynamic analysis-driven repairs. In addition, to avoid generating repetitive patches for the same underlying issue, the Patch Coder maintains a scratchpad history of previous root cause analyses and generated patches during the repair iterations.

\subsubsection{Iterative Patch Verification and Refinement} After generating a patch, Patch Coder tests it against the public test suite. If tests fail, it'll select the first failing test as the new focus and restart the debugging process. The failing test, along with the updated buggy code and error messages, goes back to Program Inspector for new dynamic analysis. This creates an iterative process where each round builds upon previous findings while fixing newly discovered issues.
The framework terminates under two conditions: (1) when the generated patch passes all test cases, Patch Coder returns the successfully repaired code, or (2) when the maximum number of patch attempts is exceeded, it returns the current code version with test results. This design ensures productive iteration while preventing infinite loops and maintaining debugging efficiency.

\subsection{InspectWare: Middleware for Reliable Debugger Integration}

Beyond action orchestration, another critical challenge in LLM-debugger interaction is operating stateful tools. Unlike stateless tools (e.g., \texttt{grep}, \texttt{sed}, web search) commonly used in existing agent systems~\cite{24sweagent,25lingmagpt}, debuggers maintain complex execution states where each command depends on and modifies the current session context. For instance, navigating to a breakpoint in Python Debugger establishes a specific execution state, and subsequent commands like \texttt{continue} or \texttt{interact} operate from this context. Our preliminary experiments revealed that LLMs struggle with direct debugger interaction, experiencing frequent state corruption and error loops (detailed in RQ3). This stateful complexity has hindered debugger adoption in the program repair community.

To bridge this gap, we developed \textbf{InspectWare}, a specialized middleware that transforms the stateful debugger into a stable, LLM-friendly interface. InspectWare not only abstracts debugger complexity but also enables high-quality process reward signals that guide LLMs toward correct debugging actions and patch hypotheses. Unlike existing debugging methods~\cite{24LDB,25AutoSD} that may propagate errors through irreversible file edits, our middleware enables immediate validation of hypotheses, preventing error accumulation and improving repair efficiency. Specifically, InspectWare provides:

\subsubsection{Stateful Session Management.} InspectWare abstracts debugger states into five clear modes (Start, Runtime State, Runtime Error, Post Mortem Mode, and Done) and continuously tracks state transitions. As shown in Figure \ref{fig:inspectcoder_framework}, this abstraction prevents invalid operations that would corrupt debugging sessions (eg. set breakpoints in post mortem mode). For each runtime state, InspectWare monitors contextual information shown in Listing \ref{fig:inspectware_status} for Inspector, maintaining LLM awareness of the current execution context and programatically preventing state-invalid commands—a critical foundation for stable debugger interactions.
\noindent \begin{center}
\begin{minipage}{0.95\columnwidth}
\begin{lstlisting}[
 basicstyle=\footnotesize\ttfamily,
 breaklines=true,
 breakindent=8pt,
 showstringspaces=false,
 rulecolor=\color{MyPurple},
 framerule=1pt,
 caption={Example of InspectWare extracting PDB state information for Program Inspector.},
 label={fig:inspectware_status}
]
## PDB Execution Status
Current Stack Trace:
[1] <module> at __test__.py:95| testcases.test_case_2()
[2] test_case_2 at __test__.py:66| result = task_func(...)
[3] task_func at __test__.py:38| if not matching_files: (paused here)
Active Breakpoints:
b1 __test__.py:38, hit 0 times.
\end{lstlisting}
\end{minipage}
\end{center}

\subsubsection{Enhanced Runtime Modification.} Since PDB's built-in \texttt{interact} session requires tedious enter/exit operations, InspectWare implements a custom interactive environment using Python's \texttt{code} library. This enables direct runtime modifications that always update to the program's execution context, providing consistent process reward signals to Inspector agent. When the Inspector tests a hypothesis, it receives immediate feedback about whether the modification resolves downstream errors—enabling rapid hypothesis refinement without costly re-execution cycles. Our implementation maintains compatibility with diverse Python test frameworks (e.g., unittest~\cite{unittest} and pytest~\cite{pytest}), including mock objects and I/O testing.

\subsubsection{Transparent Interface Design.} InspectWare provides a clean, terminal-style interface while automatically handling complex backend operations. The middleware parses LLM actions, manages state consistency, filters verbose outputs, and provides immediate feedback for invalid actions. This abstraction layer reduces context consumption while ensuring reliable meaningful process rewards. By standardizing debugger interaction, InspectWare paves the way for broader adoption of LLM-driven dynamic analysis in real-world development.

\section{Experimental Design}
\subsection{Research Questions}
Our evaluation addresses the following research questions:

\subparagraph{\textbf{RQ1 (Effectiveness):}} \textbf{\emph{How effective is InspectCoder in bugs fixing?}} We evaluate InspectCoder's bug resolving capabilities on challenging self repair benchmarks and compared to SOTA baselines.

\subparagraph{\textbf{RQ2 (Efficiency):}} \textbf{\emph{How efficient is InspectCoder in bugs fixing?}} We investigate the time and computational costs of InspectCoder's debugging process compared to SOTA baselines.

\subparagraph{\textbf{RQ3 (Ablation Study):}} \textbf{\emph{What are the key design principles for interactive debugging agents?}} We investigate which debugging capabilities and debugger-integration approaches are most effective by comparing InspectCoder variants with different design choices, revealing actionable principles for practitioners building debugger-integrated agentic systems.

\subparagraph{\textbf{RQ4 (Empirical Study):}} \textbf{\emph{What debugging patterns does InspectCoder exhibit during dynamic analysis?}} We analyze the frequency of different debugger actions (breakpoint placement, variable inspection, runtime modification) and conduct case study on the orchestrated tool call patterns to understand how InspectCoder perform interactive debugging and identify key behavioral insights.

\subparagraph{\textbf{RQ5 (Generalizability):}} \textbf{\emph{How well does InspectCoder generalize across different LLM architectures?}} We evaluate InspectCoder's performance with multiple SOTA LLMs to assess the framework's robustness and broader applicability.

\subsection{Benchmarks}
Following established practices in self-repair evaluation~\cite{24NExT,24CYCLE,25AutoSD,25KnowledgeAPR}, we construct our benchmarks from LLM-generated buggy code. Specifically, we employ Qwen2.5-Max \cite{qwen25} to generate initial code solutions on the full sets of two challenging program synthesis datasets: BigCodeBench \cite{25bigcodebench} and LiveCodeBench \cite{25livecodebench}. 
We then collect buggy solutions and filter out trivial syntactic errors (e.g., indentation errors, syntax errors) that are caught by the test framework. This leaves execution errors (e.g. TypeError) and runtime errors (e.g., assertion failures, incorrect outputs) that represent non-trivial semantic gaps from task requirements. We further filter out bugs that are only triggered by private tests, which are inaccessible during real-world development. Finally, we build two program repair benchmarks spanning diverse debugging scenarios:

\textbf{BigCodeBench-R:} BigCodeBench-R contains buggy solutions to realistic and challenging programming problems like data analysis and web development. These problems requires the interactions with multiple function calls from external modules. We obtain 607 valid \textit{<programming problem, buggy code solution>} pairs meeting the aforementioned filtering criteria from 1,140 problems in BigCodeBench, to create our BigCodeBench-R benchmark.

\textbf{LiveCodeBench-R:} LiveCodeBench-R contains buggy solutions to contamination-free competition level programming problems sourced from LeetCode, AtCoder, CodeForces. These problems typically involves more complex algorithmic logic with limited reliance on external APIs. We collect 151 valid \textit{<programming problem, buggy code solution>} pairs meeting the aforementioned filtering criteria from 279 problems in LiveCodeBench with problem release date spanning from 2024-08-01 to 2025-02-01 period, to create our LiveCodeBench-R benchmark.

\subsection{Metrics}
We employ the following metrics to evaluate the performance of InspectCoder and baseline approaches:

\textbf{Resolve Rate:} Our primary metric is the bug resolve rate, calculated as the percentage of successful fixes out of the total number of bugs. This metric directly measures the debugging \emph{effectiveness} at bridging the requirement-implementation gaps.

\textbf{Pass Rate:} We additionally report the pass rate on the full set of BigCodeBench and LiveCodeBench after applying the program repair methods. This metric shows how much these methods enhance LLMs coding capabilities beyond their initially cognitive biases.

\textbf{\#Fixes/Hour:} We record the debug time for each program repair method across all problems , then calculate the average number of bugs successfully resolved per hour to quantify time efficiency. This represents the practical \emph{efficiency} that developers experience on these methods.

\textbf{\#Fixes/Dollar:} We calculate the monetary cost per problem based on LLM API usage using $\text{Cost} = (\text{Input tokens} \times \text{Input price}) + (\text{Output tokens} \times \text{Output price})$, 
 then compute the average number of successful fixes per dollar to quantify cost efficiency.
\subsection{Baselines}
we compare InspectCoder against SOTA LLM-based program repair approaches that applies the varying debugging paradigms:

\textbf{Simple-Debugging:} This represents the most straightforward static debugging method, where LLMs are provided with no dynamic information and are asked to directly generate code patches iteratively. The process iterates until all test cases pass, representing a baseline approach without sophisticated debugging analysis.

\textbf{Rubber-Ducking \cite{24Self-Debug}:} The rubber duck debugging paradigm requires LLMs to first explain the code line-by-line, compare the explanation against the problem description, then propose patches based on identified discrepancies. This paradigm has been widely adopted in recent work~\cite{PATCH,24FixAgent,24Self-Debug} and represents the performance of prevalent static debugging methods that do not utilize dynamic execution information other than the original test error messages. 

\textbf{Trace-Reasoning \cite{24Self-Debug}:} This approach asks LLMs to verbally trace through the program's execution flow for given inputs, predicting what happens at each step without actually running the code. It represents debugging attempts that reasoning over the dynamic information using only static simulation, rather than actual program execution.

\textbf{AutoSD \cite{25AutoSD}:} This approach employs a fixed workflow for explainable debugging, where LLMs iteratively re-execute a template script to first instrument and execute code with single print statement at a time to collect additional logs before patching the buggy code. This approach essentially mimics a print-style debugging practice that  belongs to our log-augmented debugging category.

\textbf{LDB \cite{24LDB}:} This approach follows an iterative 2-step workflow that indiscriminately instruments all variable values throughout every code line, and then executes programs to collect state logs to LLMs for log-augmented debugging. We evaluate three variants on LDB's instrumentation granularities: LDB (Line-Level) traces all variables at each line, LDB (Function-Level) traces all variables before and after function calls, and LDB (Block-Level) traces all variables at each code block boundary.

\textbf{SWE-Agent \cite{24sweagent}}: We also include SWE-Agent, a well-established baseline from repository-level issue resolution. SWE-Agent employs agentic workflow with shell-based commands (e.g., sed, cat, file editing) to iteratively examine and modify code files. While it is primarily designed for repository exploration and issue localization—a complementary challenge to our focus on diagnosing complex logic errors with explicit requirements—it represents a prevalent limitation in program repair: sophisticated agentic systems that nonetheless overlook the intermediate runtime state in programs. This comparison demonstrates whether interactive debugger control provides distinct advantages beyond general agentic code manipulation. 
\subsection{Implementation Details}
Our experiments are conducted under the following settings:

\textbf{LLMs:} 
We employ SOTA LLMs to validate our approach. We use Qwen-2.5-Max \cite{qwen25} for main experiments and evaluate on DeepSeek-V3 \cite{deepseek-v3}, GPT-4o \cite{gpt4o}, and Claude-3.5-Sonnet~\cite{claude35} to demonstrate generalizability. We access all models through unified API services and calculate monetary costs based on models' official pricing. To minimize the impact of network latency overhead, we conduct experiments for all models and baselines within similar time windows under the same network environment when measuring debug time.

\textbf{Baseline Replication:}
We replicate all baselines using their publicly released code and prompts. We make minimal prompt adjustments to ensure all methods receive the same input: $D$, $C_{\text{buggy}}$, $T$. For AutoSD, we adapt the script parser to better handle edge cases and extend its patch template to support multi-line edits needed in our tasks.For SWE-Agent, we adapt the initial analysis step of "Recommended Workflow" from "Analyze the codebase by finding and reading relevant files" to "Analyze the buggy code and test case in the given code file" to align with our repair scenario, while retaining its original shell-based debugging capabilities (sed, cat, nl, etc.).

\textbf{Iteration Limits:} All methods are constrained to a maximum of 5 patch attempts. A bug is considered resolved if a plausible patch that passes all test cases is generated within these 5 rounds. For methods with variable inference steps (AutoSD and InspectCoder), we additionally impose a limit of 20 maximum reasoning iterations to ensure fair comparison and prevent excessive computation.

\textbf{Test Case and Validation:} We follow standard practices for test case usage: for LiveCodeBench-R, we use only public test information during debugging and evaluate patches against the complete public+private test suite; for BigCodeBench-R that does not differentiate the tests, we use all available test information during debugging, consistent with prior work~\cite{24NExT,25KnowledgeAPR}. Following established evaluation protocols~\cite{25KnowledgeAPR,swebench,autocoderover}, we compute resolve rates based on plausible patches that pass all test cases. Since the benchmarks do not provide reference implementations, it is almost infeasible to exhausively manually verify the strict correctness across thousands of generated patches. To validate that plausible patches reliably indicate correct repairs, we manually analyzed a random sample of 43 plausible patches by InspectCoder (10\%) and observed 43/43 semantic correctness. This high precision suggests that the comprehensive test suites—featuring extensive edge case coverage as noted by BigCodeBench \& LiveCodeBench—effectively serve as reliable correctness oracles, making plausible patches a valid evaluation proxy for successful bug fixes.

\section{Results}
\begin{table*}[th]
\centering
\caption{Comparing with Baselines: using Pass Rate metrics (on BigCodeBench) and program repair metrics (on BigCodeBench-R). BI: Breakpoint Inspection; RM: Runtime Modification.}
\begin{threeparttable}
\begin{tabular}{lcc|cccccc}
\toprule
\multicolumn{1}{c}{\textbf{Method}} & \multicolumn{1}{c}{\textbf{Pass}} & \multicolumn{1}{c|}{\textbf{$\Delta$Pass}} & \multicolumn{1}{c}{\textbf{Resolve}} & \multicolumn{1}{c}{\textbf{Debug}} & \multicolumn{1}{c}{\textbf{\#Fixes/}} & \multicolumn{1}{c}{\textbf{Debug}} & \multicolumn{1}{c}{\textbf{\#Fixes/}} \\
\multicolumn{1}{c}{} & \multicolumn{1}{c}{\textbf{Rate}} & \multicolumn{1}{c|}{\textbf{Rate}} & \multicolumn{1}{c}{\textbf{Rate*}} & \multicolumn{1}{c}{\textbf{Time}} & \multicolumn{1}{c}{\textbf{Hour}} & \multicolumn{1}{c}{\textbf{Cost}} & \multicolumn{1}{c}{\textbf{Dollar}} \\
\hline
\rowcolor{gray!20}
One-pass generation & 45.26\% & +0\% & 0\% & 0s & 0 & \$0 & 0 \\
Simple-Debugging & 49.91\% & +4.65\% & 8.73\% & \underline{146.40s} & 2.15 & \$0.0084 & 10.39 \\
Rubber-Ducking & 75.61\% & +30.35\% & 57.00\% & 208.93s & 9.82 & \$0.0179 & 31.84\\
Trace-Reasoning & 76.31\% & +31.05\% & 58.32\% & 202.14s & 10.39 & \$0.0198 & 29.45 \\
SWE-Agent & 72.63\% & +27.37\% & 51.40\% & 388.11s & 4.77 & \$0.0333 & 15.44 \\
AutoSD& 63.51\% & +18.24\% & 34.27\% & 645.66s & 1.91 & \$0.0689 & 4.97 \\
LDB (Line-Level)& 78.92\% & +33.68\% & 63.26\% & 218.43s & 10.42 & \$0.0224 & 28.24 \\
LDB (Block-Level)& \underline{79.65\%} & \underline{+34.39\%} & \underline{64.58\%} & 194.30s & \underline{11.97} & \$0.0227 & 28.45 \\
LDB (Function-Level)& 78.24\% & +32.98\% & 61.94\% & 203.33s & 10.96 & \$0.0235 & 26.36 \\
\midrule
\textbf{InspectCoder} & \textbf{81.40\%} & \textbf{+36.14\%} & \textbf{67.87\%} & \textbf{121.74s} & \textbf{20.07} & \$0.0215 & 31.57 \\
w/o BI Strategy & 77.80\%& +32.54\%& 61.12\%& 206.53s& 10.65 & \$0.0244 & 25.05 \\
w/o RM Strategy & 80.25\%& +34.99\%& 63.92\%& 181.63s& 12.67 & \$0.0197 & 32.45 \\
w/o debugger tool& 76.61\%& +31.34\% & 58.86\% & 558.86s& 3.80 & \$0.0221 & 26.64 \\
w/o InspectWare & 58.94\% & +13.68\% & 25.70\% & 235.08s & 2.85 & \$0.0313 & 8.21 \\

\bottomrule
\end{tabular}
\begin{tablenotes}
\small
\item[*] Resolve Rate is the primary evaluation metric
\end{tablenotes}
\end{threeparttable}
\label{tab:bcb_main_results}
\end{table*}
This section introduces the experimental results as well as the analysis and findings for each RQ.
\subsection{RQ1: Effectiveness of InspectCoder}
To answer RQ1, we present the main evaluation results regarding InspectCoder's performance in program repair. We report the bug resolve rate for Qwen2.5-Max on BigCodeBench-R and LiveCodeBench-R across different program repair methods. We additionally report the full-set pass rate to demonstrate the overall task solving  performance.
As shown in Tables \ref{tab:bcb_main_results} and \ref{tab:lcb_main_results}, the base LLM initially achieves 45.26\% and 36.20\% pass rates on BigCodeBench and LiveCodeBench respectively. On the remaining 607 and 151 challenging tasks where the LLM fails, InspectCoder successfully resolves 412 and 19 bugs respectively, substantially outperforming all baselines and bringing considerable pass rate improvements of 36.14\% and 6.81\% to the base LLM.
\begin{table*}[tbh]
\centering
\caption{Comparing with Baselines: using Pass Rate metrics (on LiveCodeBench) and prorgam repair metrics (on LiveCodeBench-R). BI: Breakpoint Inspection; RM: Runtime Modification.}
\begin{threeparttable}
\begin{tabular}{lcc|cccccc}
\toprule
\multicolumn{1}{c}{\textbf{Method}} & \multicolumn{1}{c}{\textbf{Pass}} & \multicolumn{1}{c|}{\textbf{$\Delta$Pass}} & \multicolumn{1}{c}{\textbf{Resolve}} & \multicolumn{1}{c}{\textbf{Debug}} & \multicolumn{1}{c}{\textbf{\#Fixes/}} & \multicolumn{1}{c}{\textbf{Debug}} & \multicolumn{1}{c}{\textbf{\#Fixes/}} \\
\multicolumn{1}{c}{} & \multicolumn{1}{c}{\textbf{Rate}} & \multicolumn{1}{c|}{\textbf{Rate}} & \multicolumn{1}{c}{\textbf{Rate*}} & \multicolumn{1}{c}{\textbf{Time}} & \multicolumn{1}{c}{\textbf{Hour}} & \multicolumn{1}{c}{\textbf{Cost}} & \multicolumn{1}{c}{\textbf{Dollar}} \\
\hline
\rowcolor{gray!20}
One-pass generation & 36.20\% & +0\% & 0\%& 0 & 0 & \$0 & 0 \\
Simple-Debugging & 37.88\% & +1.68\% & 3.31\%& \textbf{194.44s} & 0.61 & \$0.0078 & 4.24 \\
Rubber-Ducking & 39.07\% & +2.87\% & 5.30\% & 350.22s & 0.55 & \$0.0179 & 2.96 \\
Trace-Reasoning & 39.78\% & +3.58\% & 6.62\% & 377.36s & 0.59 & \$0.0198 & 3.13 \\
SWE-Agent & 39.07\% & +2.87\% & 5.30\% & 347.70s & 0.55 & \$0.0461 & 1.15 \\
AutoSD & 38.71\% & +2.51\% & 4.64\% & 518.16s & 0.32 & \$0.1996 & 0.23 \\
LDB (Line-Level) & 39.78\% & +3.58\% & 6.62\% & 396.15s & 0.60 & \$0.0447 & 1.48 \\
LDB (Block-Level) & \underline{40.50\%} & \underline{+4.30\%} & \underline{7.95\%} & 320.24s & \underline{0.89} & \$0.0446 & 1.78 \\
LDB (Function-Level) & 40.14\% & +3.94\% & 7.28\% & 313.57s & 0.83 & \$0.0386 & 1.89 \\
\midrule
\textbf{InspectCoder} & \textbf{43.01\%} & \textbf{+6.81\%} & \textbf{12.58\%} & \underline{243.89s} & \textbf{1.88} & \$0.0374 & 3.41 \\
w/o BI Strategy & 41.22\% & +5.02\% & 9.27\% & 270.44s & 1.23 & \$0.0352 & 2.63 \\
w/o RM Strategy & 42.29\% & +6.09\% & 11.25\% & 300.40s & 1.35 & \$0.0342 & 3.29 \\
w/o debugger tool & 40.50\% & +4.30\% & 7.95\% & 1057.78s & 0.27 & \$0.0348 & 2.28 \\
w/o InspectWare & 38.42\% & +2.22\% & 4.10\% & 253.27s & 0.58 & \$0.0393 & 1.04 \\
\bottomrule
\end{tabular}
\begin{tablenotes}
\small
\item[*] Resolve Rate is the primary evaluation metric
\end{tablenotes}
\end{threeparttable}
\label{tab:lcb_main_results}
\end{table*}
\paragraph{Finding 1-1: Dynamic Information is Crucial for Program Repair Success.}
We observe a clear performance hierarchy: Trace-Reasoning achieves substantial improvement over static methods (Simple Debugging, Rubber-Ducking, and SWE-Agent) by simulating program execution traces, while LDB and InspectCoder empowered with actual program execution achieve even greater improvements. This demonstrates that dynamic information is valuable for bug diagnosis.

\paragraph{Finding 1-2: Selective Information Gathering and Reversible Exploration are Key to Effective Dynamic Analysis.}
While dynamic information is essential, effectively analyzing it presents the next challenge. Importantly, the unstable performance of LDB across benchmarks under different predefined instrumentation granularities reveals a fundamental challenge for SOP-styled log-augmented debugging: differentiating useful vs. redundant dynamic information for selective collection. InspectCoder addresses this challenge by giving LLMs \textbf{the initiative to select dynamic information} through flexible debugger interactions, 

achieving 5.10\% and 60.37\% relative resolve rate increases over the best performance in LDB series.

Beyond information selection, two additional factors critically impact dynamic analysis effectiveness: \textbf{(1) the depth of analysis} and \textbf{(2) the reversibility of exploration}. File-editing approaches exemplify these challenges. AutoSD, despite accessing execution output from templated debugging scripts, shows only modest resolve rate improvements over simple static methods—revealing that shallow execute-then-conclude workflows constrain in-depth root cause analysis. Moreover, these approaches face fundamental reversibility issues: once code is modified, recovering from incorrect hypotheses becomes costly. Through empirical analysis, we observe that SWE-Agent's erroneous \texttt{sed} operations often corrupt code files (especially during multi-line replacements), forcing the agent to waste subsequent reasoning steps on file state restoration rather than bug analysis. In contrast, InspectCoder enables \textbf{multi-step exploration and analysis at breakpoints} for deep understanding, achieving 98.04\% and 171.12\% relative increases over AutoSD, while supporting instant recovery from failed hypotheses through debugger's interactive mode or restart—treating errors as valuable process reward signals rather than permanent setbacks (as shown in RQ4 case study, Pattern 3).

\paragraph{Finding 1-3: Interactive Debugging Enables Both Unique Problem-Solving and Broad Applicability.}

\begin{figure}[htbp]
  \centering
  \includegraphics[width=0.8\linewidth]{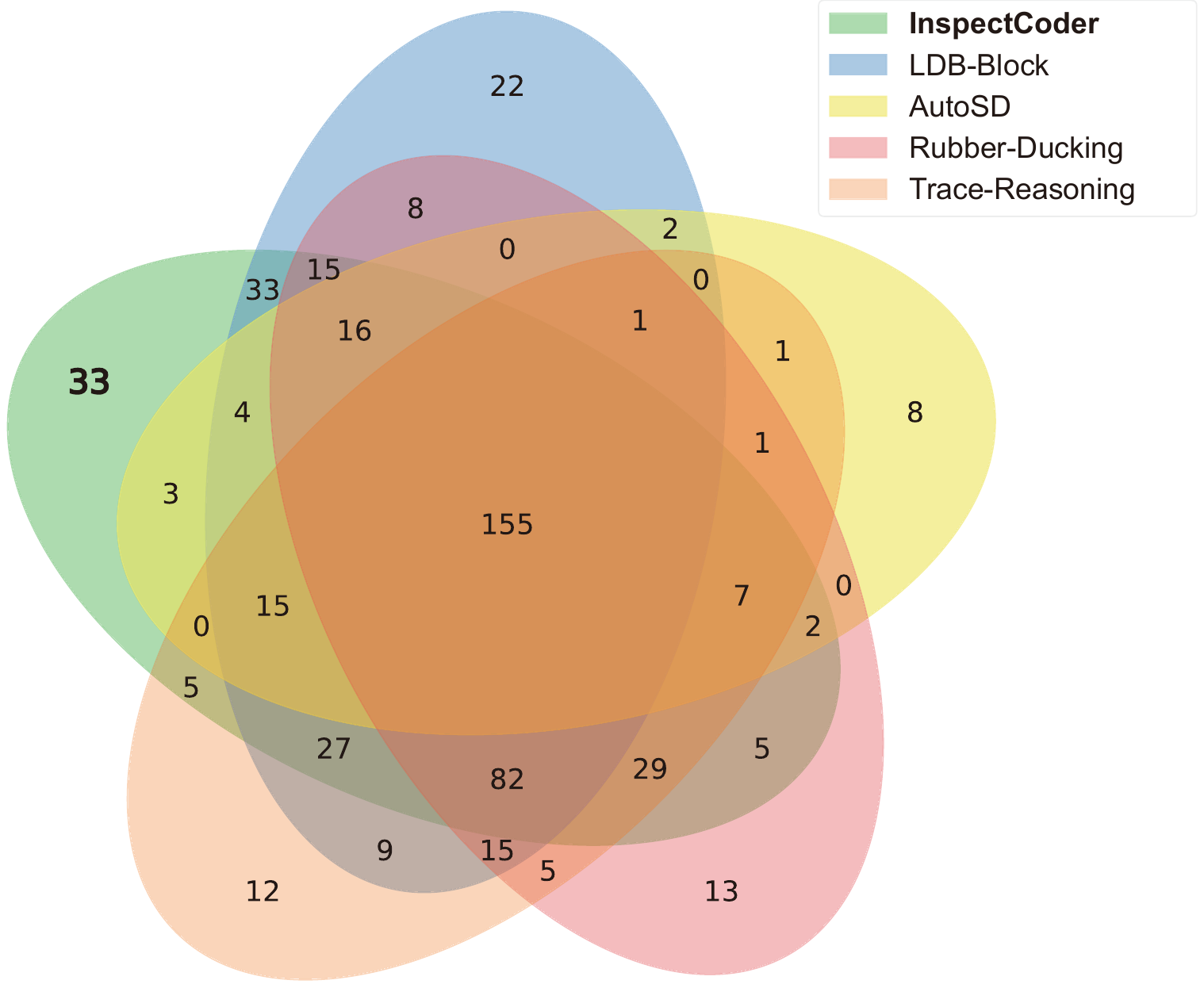}
  \caption{Bug fixes overlaps among competitive self repair methods.}
  \label{fig:intersection}
\end{figure} 
We further analyze the complementary nature of the five most competitive program repair approaches on both benchmarks in Figure \ref{fig:intersection}. InspectCoder demonstrates superior performance in both dimensions: it uniquely resolves most bugs (33 fixes) that other methods cannot handle, while also achieving the highest success rate on other-solvable bugs (10.22\% above average). This demonstrates that InspectCoder not only handles challenging cases uniquely but also performs consistently well across diverse bug types.

\begin{mybox}
  \small
  \textbf{Answer to RQ1:}
InspectCoder significantly outperforms existing methods, achieving 5.10\%-60.37\% relative improvements over the strongest baseline. Its success demonstrates that effective dynamic analysis hinges on three key factors: selective information gathering, deep multi-step exploration, and reversible debugging, which together enable both unique problem-solving capabilities and broad applicability.
\end{mybox}

\subsection{RQ2: Efficiency of InspectCoder}

In this section, we evaluate the efficiency of InspectCoder using \#Fixes/Hour, which reflects the practical deployment experience of program repair methods. As shown in Tables \ref{tab:bcb_main_results} and \ref{tab:lcb_main_results}, InspectCoder effectively controls the debug time overhead and achieves the highest time efficiency across both benchmarks, solving 1.67x and 2.24x of bugs per hour comparing with best-performing baseline LDB (Block-Level).

\begin{figure}[htbp]
  \centering
  \includegraphics[width=0.7\linewidth]{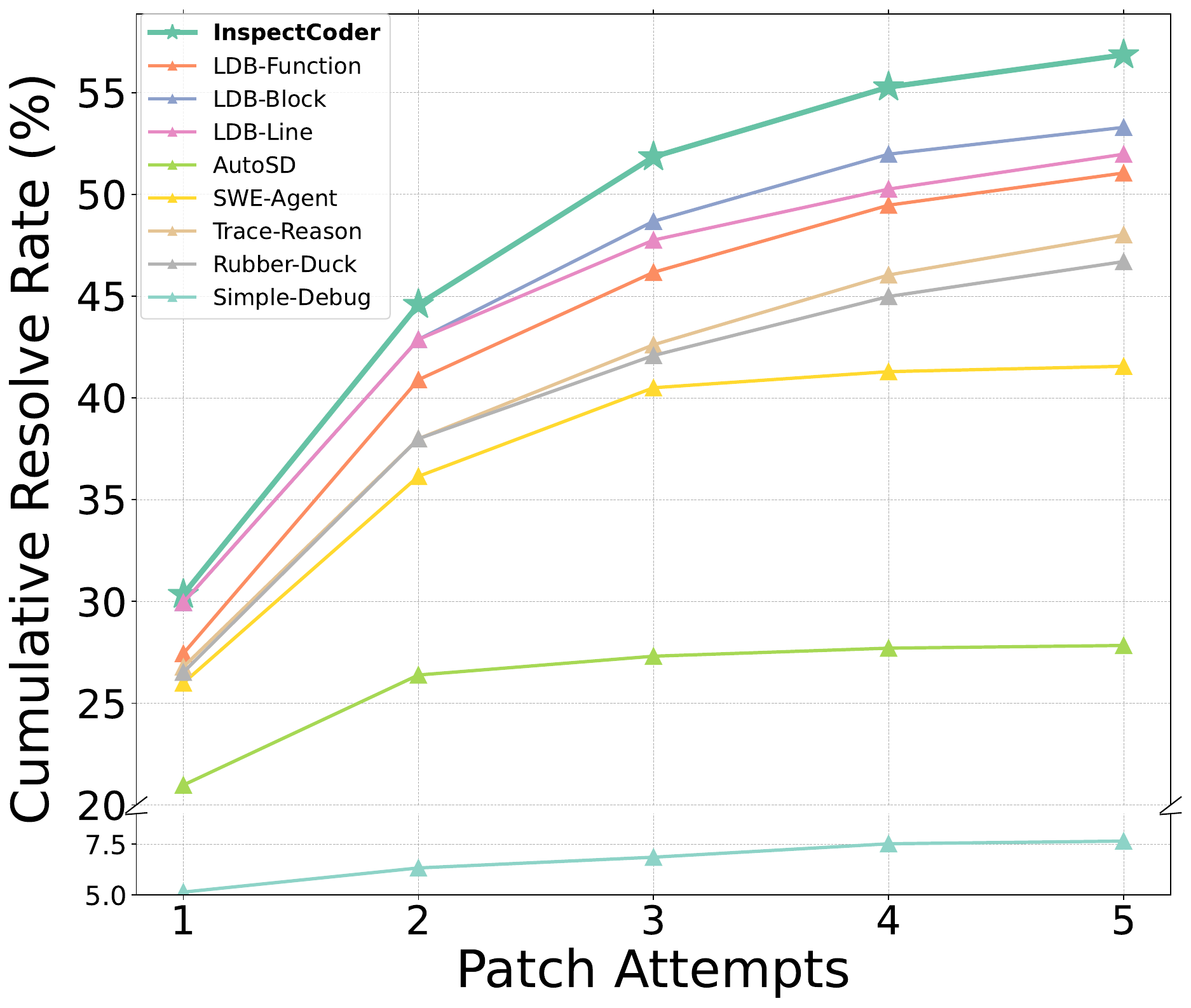}
  \caption{Impact of \#patch generation attempts on the cumulative resolve rate of InspectCoder and other baselines.}
  \label{fig:patch_attempts}
\end{figure}

InspectCoder's efficiency stems from two key aspects. \textbf{(1) Fewer patch attempts.} As shown in Figure \ref{fig:patch_attempts}, InspectCoder exhibits a steeper resolve rate curve than competitive baselines. Compared to the final resolve rate of baselines after 5 attempts, InspectCoder already outperforms Simple-Debugging and AutoSD at its first attempt, and matches SWEAgent, Trace-Reasoning and LDB series within just 2-3 attempts. \textbf{(2) Low-overhead dynamic analysis.} Unlike prior dynamic methods (AutoSD, LDB) that must instrument and re-execute the entire program 1-2 times to gather new information, InspectCoder interacts with a single live debugging session, where each reasoning step involves only brief action calls or few-line debug code execution with immediate feedback. Additionally, InspectCoder maintains manageable monetary costs (typically <5 cents per repair), ranking second in overall cost efficiency across both benchmarks. This balance of time and cost efficiency underscores its practicality for real-world deployment.

\begin{mybox}
  \small
  \textbf{Answer to RQ2:}
InspectCoder achieves the highest time efficiency across both benchmarks, solving 1.67x and 2.24x more bugs per hour than the strongest baselines. This efficiency results from fewer required patch attempts and low-overhead dynamic analysis through live debugging sessions at manageable costs.
\end{mybox}

\subsection{RQ3: Design Principles for Agentic Dynamic Analysis System}
We investigate which debugging capabilities and integration approaches are most effective by comparing InspectCoder variants with different design choices, revealing key principles for practitioners designing debugger-integrated agentic systems.
\paragraph{Insight 3-1: State Inspection is Fundamental for Dynamic Debugging.} We evaluate a variant masking out the breakpoint inspection actions (\textbf{w/o BI Strategy}), allowing only perturbations. Results show severe degradation: resolve rates drop by 6.75\% \& 3.31\%. This reveals direct observation of runtime states is the foundation of effective dynamic debugging. Practitioners should prioritize state inspection capabilities over modification features.

\paragraph{Insight 3-2: Runtime Modification Complements Inspection for Hypothesis Testing.} 
Removing runtime modification capabilities (\textbf{w/o RM Strategy}) causes performance drops of 3.95\% \& 1.33\% respectively. The more significant impact on BigCodeBench-R than LiveCodeBench-R reveals an important pattern: runtime modification proves most valuable when debugging involves complex objects and function calls where behavior is difficult to predict from state inspection alone. This suggests we should view inspection and modification as complementary capabilities rather than alternatives.

\paragraph{Insight 3-3: Actual Debugger Integration is Essential.} Inspired by Trace-Reasoning \cite{24Self-Debug} and recent LLM code reasoning studies \cite{24CruxEval,25codesense}, we investigate whether LLM simulation can replace actual debuggers (\textbf{w/o debugger tool}), where an LLM agent simulates debugger outputs. Results show severe degradation: resolve rates drop by 8.01\% \& 4.64\%, with debug time increasing dramatically (437s \& 813.89s). This demonstrates that current LLMs cannot reliably simulate dynamic program behavior, making actual debugger integration indispensable. Practitioners should integrate real debugging tools rather than relying on LLM reasoning.

\paragraph{Insight 3-4: Middleware Abstraction is Critical for Stateful Tool Integration.} 
In our early development, we attempted direct LLM-debugger interaction (\textbf{w/o InspectWare}), which revealed two critical failure modes: (1) LLMs fail to maintain stateful interactions when issuing invalid commands (e.g., setting breakpoints in post-mortem mode), leading to session corruption; (2) Verbose, debugger-specific outputs confuse and overwhelm LLMs. Our middleware design addresses these issues through state abstraction and output filtering and rewriting, improving performance by 5.44\% \& 2.65\%. This demonstrates that middleware layers are essential for bridging LLMs with stateful tools. Practitioners should implement abstraction layers for robust state management and output formatting. Looking forward, enhancing LLMs' native debugger reasoning through targeted training could benefit both debugging and broader code understanding tasks.

\begin{mybox}
  \small
  \textbf{Answer to RQ3:}
Our study reveals four key design principles for debugger-integrated agentic systems: (1) \textbf{prioritize state inspection} as the foundation for dynamic debugging; (2) \textbf{enable runtime modification} for hypothesis testing in complex scenarios; (3) \textbf{integrate actual debuggers} rather than LLM simulation; (4) \textbf{implement middleware abstraction} for robust stateful tool management. These principles provide actionable guidance for practitioners building LLM-debugger systems.
\end{mybox}

\subsection{RQ4: Dynamic Analysis Behaviors in InspectCoder}
\begin{figure}[htbp]
  \centering
  \includegraphics[width=\linewidth]{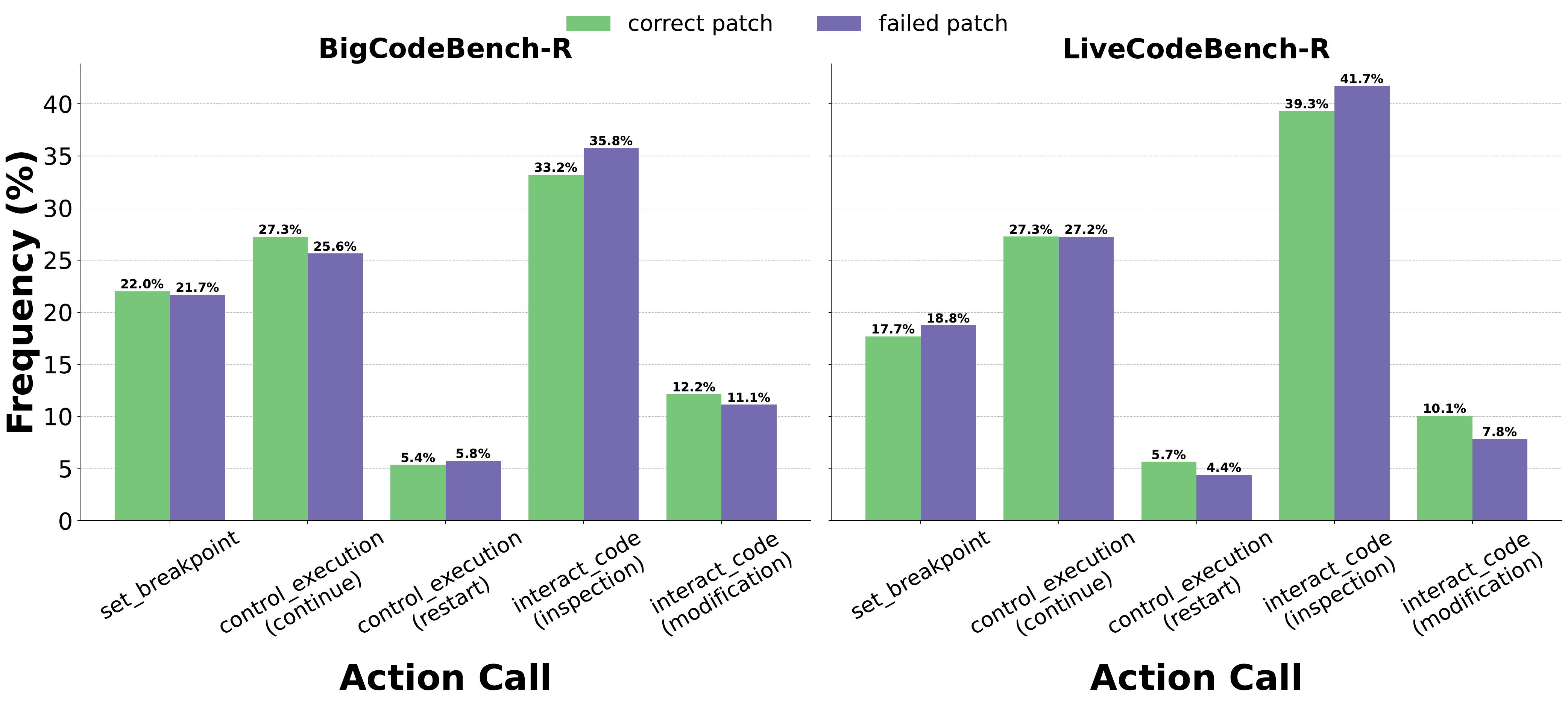}
  \caption{Frequency of action Calls during Dynamic Analysis.}
  \label{fig:action_frequency}
\end{figure}

\subsubsection{Quantitative Analysis.} To understand InspectCoder's debugging behavior, we analyze how the Program Inspector utilizes different actions during dynamic analysis. On average, InspectCoder performs 32.76 (BigCodeBench-R) and 71.77 (LiveCodeBench-R) interaction actions per task.

Figure \ref{fig:action_frequency} shows the frequency distribution of different action types. The most frequently used action is \texttt{interact\_code}, which serves as our core mechanism for dynamic analysis, averaging 15.42 calls (47.0\%) on BigCodeBench-R and 36.34 calls (50.6\%) on LiveCodeBench-R, with breakpoints set 6.52 and 12.87 times per task respectively.

We categorize \texttt{interact\_code} into inspection and modification based on whether the debug code contains print statements. Breakpoint inspection dominates usage patterns (34.5\% on BigCodeBench-R, 40.5\% on LiveCodeBench-R), confirming its role as the primary method for gathering runtime information. In contrast, runtime modification shows more conservative usage (11.5\% and 8.95\% respectively).
Interestingly, we observe contrasting patterns between successful and failed cases across both datasets. In tasks that result in correct patches, LLMs use modification more frequently while relying less on inspection. Conversely, in failed cases, LLMs perform extensive inspection but remain conservative with modifications. This suggests that on challenging tasks, LLMs struggle to understand the root cause and continuously inspect runtime states for deeper comprehension, while being hesitant to make experimental changes due to incomplete bug understanding. In contrast, when LLMs have sufficient confidence from their analysis, they are more willing to use runtime modification to validate debugging hypotheses. We further conduct case studies to examine how exactly does InspectCoder debug compairing with baselines.

\subsubsection{Case Study.}
While the quantitative analysis reveals the overall frequency of debugger actions, it does not capture how InspectCoder strategically combines these actions to form effective debugging patterns. To provide deeper insights into InspectCoder's debugging behavior, we examine representative debugging sessions that showcase distinct patterns emerged during our experiments. These patterns illustrate how InspectCoder leverages its dynamic analysis capabilities to overcome limitations of static debugging approaches.

We identify three recurring debugging patterns that emerges in InspectCoder's sophisticated use of interactive debugging:

\paragraph{Pattern 1: Multi-Hop State Tracing.}
This pattern demonstrates InspectCoder's ability to systematically trace variable evolution through program execution. In the example shown, the Inspector tracks how duplicate column values flow through multiple processing stages (shuffling and sorting operations) by strategically placing breakpoints at critical junctures. This multi-hop tracing reveals that duplicates persist throughout the pipeline—an insight difficult to obtain through static analysis or single-point logging. While log-based approaches like LDB would instrument all variables indiscriminately, generating overwhelming output, InspectCoder selectively inspects only the relevant variables at precise execution points, maintaining focus on the bug's causal chain.
\begin{tcolorbox}[
    title={Pattern 1: Multi-Hop State Tracing}, 
    colback=white,
    colframe=gray!75,
    fonttitle=\bfseries
]

\textbf{Bug Context:} DataFrame creation with duplicate columns causing test failure

\tcblower

\fcolorbox{blue!30}{blue!5}{\parbox{0.97\linewidth}{
\textbf{Thought:} \textit{The breakpoint at line 25 failed (blank line). Need to set it at line 26 where columns are shuffled. After pausing, I'll inspect the \texttt{columns} list to understand duplicate handling.}
}}

\vspace{0.3em}
\fcolorbox{green!30}{green!5}{\parbox{0.97\linewidth}{
\textbf{Action \& Observation:}\\[0.2em]
\texttt{> set\_breakpoint(26) -> control\_execution('continue')}\\
\texttt{← Paused before line 26 at test.py}\\
\texttt{> interact\_code('print(columns)')}\\
\texttt{← ['A', 'A', 'B', 'B', 'C']}
}}

\vspace{0.5em}
\hrule
\vspace{0.5em}

\fcolorbox{blue!30}{blue!5}{\parbox{0.97\linewidth}{
\textbf{Thought:} \textit{The \texttt{columns} list contains duplicates as expected. Now I need to trace how these duplicates flow through the shuffling and sorting operations. Setting breakpoint at line 29 to inspect both \texttt{shuffled\_columns} and the sorting process.}
}}

\vspace{0.3em}
\fcolorbox{green!30}{green!5}{\parbox{0.97\linewidth}{
\textbf{Action \& Observation:}\\[0.2em]
\texttt{> set\_breakpoint(29) -> control\_execution('continue')}\\
\texttt{← Paused before line 29 at test.py}\\
\texttt{> interact\_code('print(shuffled\_columns)')}\\
\texttt{← ['C', 'B', 'B', 'A', 'A']}
}}

\vspace{0.5em}
\textbf{Key Finding:} Through multi-hop tracing, Inspector identifies that duplicates persist through the processing pipeline, requiring further investigation into DataFrame construction.
\end{tcolorbox}

\paragraph{Pattern 2: Inspect-Perturb-Validate Loop.}
This pattern showcases the seamless integration of passive observation and active experimentation. The Inspector first inspects a complex data structure (nested tuple) to understand the bug's nature, then immediately perturbs the runtime state to test a fix hypothesis. The debugger provides instant process reward—program successful completion—confirming the solution's validity without requiring a full patch-test cycle. This rapid feedback loop is particularly powerful: what would require multiple iterations of code modification, file saving, and test re-execution in approaches like AutoSD or SWE-Agent is accomplished in a single debugging session. The pattern exemplifies why successful cases in our quantitative analysis show higher runtime modification usage—confident understanding enables decisive action.
\begin{tcolorbox}[
    title={Pattern 2: Inspect-Perturb-Validate Loop}, 
    colback=white,
    colframe=gray!75,
    fonttitle=\bfseries
]

\textbf{Bug Context:} TypeError when converting tuple elements to integers with \texttt{map(int, T1)}

\tcblower

\fcolorbox{blue!30}{blue!5}{\parbox{0.97\linewidth}{
\textbf{Thought:} \textit{Set breakpoint at line 6 where \texttt{map} function is called to inspect \texttt{T1} contents before the error occurs.}
}}

\vspace{0.3em}
\fcolorbox{green!30}{green!5}{\parbox{0.97\linewidth}{
\textbf{Action \& Observation:}\\[0.2em]
\texttt{> set\_breakpoint(6) -> control\_execution('continue')}\\
\texttt{← Paused before line 6 at test.py}\\
\texttt{> interact\_code('print(T1)')}\\
\texttt{← (('1', '2', '3'),)}
}}

\vspace{0.5em}
\hrule
\vspace{0.5em}

\fcolorbox{blue!30}{blue!5}{\parbox{0.97\linewidth}{
\textbf{Thought:} \textit{\texttt{T1} is a nested tuple structure. The \texttt{map(int, T1)} is trying to convert the inner tuple itself rather than its elements, causing TypeError. I'll flatten \texttt{T1} using \texttt{itertools.chain} and validate this fixes the issue.}
}}

\vspace{0.3em}
\fcolorbox{green!30}{green!5}{\parbox{0.97\linewidth}{
\textbf{Action \& Observation:}\\[0.2em]
\texttt{> interact\_code('T1 = list(itertools.chain(*T1))')}\\
\texttt{← (Successfully run. No output)}\\
\texttt{> control\_execution('continue')}\\
\texttt{← Program execution finished successfully and automatically restarted.}
}}

\vspace{0.5em}
\hrule
\vspace{0.5em}

\fcolorbox{blue!30}{blue!5}{\parbox{0.97\linewidth}{
\textbf{Thought:} \textit{Perfect! This confirms my hypothesis was correct. Now I'll propose the repair.}
}}

\vspace{0.3em}
\fcolorbox{green!30}{green!5}{\parbox{0.97\linewidth}{
\textbf{Action:}\\[0.2em]
\texttt{> propose\_repair('The root cause of the bug is that T1 is a tuple of tuples, and the map(int, T1) function is trying to convert the inner tuple to integers, which is not possible. To fix this, we need to flatten T1 before  passing it to the map function...)}
}}

\vspace{0.5em}
\textbf{Key Finding:} InspectCoder seamlessly combines inspection of complex data structures with immediate runtime perturbation, receiving instant process reward signals without costly re-execution—a debugging cycle impossible with static analysis or log-based approaches.
\end{tcolorbox}

\paragraph{Pattern 3: Resilient Error Recovery.}
Not all debugging hypotheses prove correct, and this pattern illustrates InspectCoder's graceful handling of failed attempts. When a runtime modification triggers an assertion error and enters post-mortem mode, the Inspector treats this negative signal as valuable feedback rather than a setback. Through simple restart, it instantly recovers to a clean state and adjusts its investigation strategy, ultimately discovering the deeper issue (missing dtype specifications). In contrast, file-editing approaches like SWE-Agent would have already committed potentially incorrect changes to the codebase, requiring extra effort in identification and reversal of problematic modifications. This reversibility transforms debugging from a cautious, linear process into an exploratory one where failed attempts contribute to understanding rather than creating technical debt.
\begin{tcolorbox}[
    title={Pattern 3: Resilient Error Recovery}, 
    colback=white,
    colframe=gray!75,
    fonttitle=\bfseries
]

\textbf{Bug Context:} Empty DataFrame missing column definitions, shape (0,0) instead of (0,2)


\fcolorbox{blue!30}{blue!5}{\parbox{0.97\linewidth}{
\textbf{Thought:} \textit{DataFrame shape is (0,0) but needs columns ['File', 'Size']. Pandas doesn't infer columns from empty lists. Let me explicitly set column names.}
}}

\vspace{0.3em}
\fcolorbox{green!30}{green!5}{\parbox{0.97\linewidth}{
\textbf{Action \& Observation:}\\[0.2em]
\texttt{> interact\_code('df=pd.DataFrame(file\_info, columns=[...]); print(df)')}\\
\texttt{← Empty DataFrame Columns: [File, Size]}\\
\texttt{> control\_execution('continue')}\\
\texttt{← AssertionError. PDB session entering post mortem mode at ...}
}}

\vspace{0.5em}
\hrule
\vspace{0.5em}

\fcolorbox{blue!30}{blue!5}{\parbox{0.97\linewidth}{
\textbf{Thought:} \textit{Assertion failed—debugger feedback suggests incomplete fix. Might be dtype issues. Restarting to investigate more carefully.}
}}

\vspace{0.3em}
\fcolorbox{green!30}{green!5}{\parbox{0.97\linewidth}{
\textbf{Action \& Observation:}\\[0.2em]
\texttt{> control\_execution('restart')}\\
\texttt{← Post mortem finished, the test will be restarted}\\
\texttt{> set\_breakpoint(85) -> control\_execution('continue')}\\
\texttt{← Paused before line 85 at test.py}\\
\texttt{> interact\_code('print(expected\_df.dtypes)')}\\
\texttt{← File    object     Size    object}
}}

\vspace{0.5em}
\textbf{Key Finding:} Interaction errors become valuable process reward signals. InspectCoder instantly reverts failed modifications through debugger, while file-editing approaches like SWE-Agent leave behind corrupted code requiring tedious and error-prone cleanup steps.
\end{tcolorbox}

\paragraph{Synthesis.}
These patterns reveal how InspectCoder's interactive debugging transcends the limitations of static analysis and log-based approaches. The ability to flexibly combine inspection and perturbation, coupled with immediate process rewards and reversible exploration, enables a fundamentally different debugging paradigm. While static methods must reason about program behavior abstractly and log-based approaches drown in verbose output, InspectCoder maintains focused, hypothesis-driven investigation with real-time validation. This explains the efficiency gains observed in RQ2 and the behavioral differences in action usage between successful and failed cases—effective debugging requires not just information access, but the ability to interactively explore, test, and recover from incorrect hypotheses.

\begin{mybox}
  \small
  \textbf{Answer to RQ4:}
InspectCoder shows adaptive debugging patterns with \texttt{interact\_code} comprising ~50\% of interactions, dominated by breakpoint inspection (34.5-40.5\%) over runtime modification (8.95-11.5\%). Successful cases involve more modification and less inspection, while failed cases show the opposite pattern, indicating that LLMs adapt their strategies based on bug comprehension—exploring extensively when uncertain but experimenting confidently when understanding is sufficient.
\end{mybox}

\begin{table}[ht]
\centering
\caption{Comparing with Representative Baselines across Different Models on LiveCodebench-R}
\resizebox{\textwidth}{!}{
\begin{tabular}{l cc cc cc cc}
\toprule
\multirow{2}{*}{\textbf{Method}} & \multicolumn{2}{c}{\textbf{Qwen2.5-Max}} & \multicolumn{2}{c}{\textbf{DeepSeek-V3}} & \multicolumn{2}{c}{\textbf{GPT-4o}} & \multicolumn{2}{c}{\textbf{Claude-3.5}} \\
\cline{2-9}
& \textbf{Resolve} & \textbf{\#Fixes/} & \textbf{Resolve} & \textbf{\#Fixes/} & \textbf{Resolve} & \textbf{\#Fixes/} & \textbf{Resolve} & \textbf{\#Fixes/} \\
& \textbf{Rate} & \textbf{Hour} & \textbf{Rate} & \textbf{Hour} & \textbf{Rate} & \textbf{Hour} & \textbf{Rate} & \textbf{Hour} \\
\midrule
Simple-Debugging & 3.31\% & 0.61 & 5.96\% & 1.63& 3.31\% & \underline{2.25}& 5.96\% & 2.32 \\
Rubber-Ducking & 5.30\% & 0.55 & 6.62\% & 0.70 & 3.31\% & 0.79 & 7.28\% & 0.99 \\
Trace-Reasoning & 6.67\% & 0.64 & 10.67\% & 1.02 & \underline{6.62\%}& 1.55 & 8.61\% & 1.58 \\
AutoSD & 4.64\% & 0.23 & 13.04\% & 0.88 & 4.03\% & 0.39 & - & - \\
LDB (Block-Level) & \underline{7.95\%}& \underline{0.89}& \underline{17.88\%}& \underline{2.95} & 4.96\%& 0.83 & \underline{17.22\%}& \textbf{3.05} \\
\textbf{InspectCoder} & \textbf{12.58\%} & \textbf{1.86} & \textbf{24.50\%} & \textbf{3.63} & \textbf{7.28\%} & \textbf{2.29} & \textbf{19.88\%} & \underline{2.81} \\
\bottomrule
\end{tabular}
\label{tab:multi_model_results}
}
\end{table}
\subsection{RQ5: Generalizability of InspectCoder}

To evaluate InspectCoder's generalization capability, we conduct experiments across three additional SOTA LLMs with varying architectures beyond Qwen2.5-Max: DeepSeek-V3, GPT-4o, and Claude-3.5-Sonnet.
Table \ref{tab:multi_model_results} shows that InspectCoder consistently achieves the highest resolve rates across all LLMs, with substantial improvements ranging from 3.97\% to 17.88\% over established baselines. Moreover, InspectCoder also maintains top-tier time efficiency on all of the four models.

Interestingly, we observe significant variations in dynamic analysis capabilities across different LLMs. GPT-4o shows limited effectiveness with dynamic information—while achieving similar performance to Qwen2.5-Max on static debugging methods, it exhibits inferior performance on dynamic debugging approaches (InspectCoder, LDB, AutoSD). In contrast, DeepSeek-V3 and Claude-3.5 demonstrate strong dynamic reasoning abilities, with significantly enhanced performance across all dynamic debugging methods and nearly double GPT-4o's InspectCoder performance. These variations suggest that the community could benefit from prioritizing dynamic program understanding during LLM pre-training, highlighting opportunities for future dynamic-aware training methodologies.

\begin{mybox}
  \small
  \textbf{Answer to RQ5:}
  InspectCoder demonstrates strong generalizability, consistently achieving the highest resolve rates across all four tested LLMs. Interestingly, models exhibit varying dynamic analysis capabilities, suggesting opportunities for enhancing dynamic reasoning capabilities in future LLM training.
\end{mybox}

\section{Discussion and Threats to Validity}
\subsection{Threats to Validity}

\subsubsection{Internal Validity}

\paragraph{Benchmark Selection and Data Leakage Mitigation.}
LLMs trained on extensive web-crawled corpora may exhibit inflated performance due to exposure to similar problems during pre-training. To mitigate this threat, we employ two state-of-the-art datasets specifically designed with contamination-aware methodologies: BigCodeBench~\cite{25bigcodebench} constructs new programming tasks from scratch using ODEX seed examples, with validation studies confirming zero contamination~\cite{contamination}. LiveCodeBench~\cite{25livecodebench} maintains data integrity through continuous updates from competitive programming platforms, with our LiveCodeBench-R utilizing problems from August 2024, establishing a temporal boundary that postdates the knowledge cutoff of all evaluated models. Both benchmarks provide comprehensive test suites that enable rigorous evaluation of our interactive debugging capabilities while representing diverse programming challenges including algorithmic problems, data structure manipulations, and real-world utility functions. This controlled setting with well-defined specifications isolates our core contribution from confounding factors like incomplete requirements or ambiguous test cases, providing a strong foundation for validity.

\subsubsection{External Validity}

\paragraph{Generalizability Across Programming Languages.}
Our current implementation focuses on Python and leverages Python-specific debugger tools (PDB). However, the underlying paradigm—empowering LLMs with interactive debugging capabilities—is fundamentally language-agnostic. Other mainstream languages provide comparable debugging infrastructure: Java offers JDB, C/C++ provides GDB, JavaScript supports V8 Inspector Protocol, and Go includes Delve debugger. These tools support essential capabilities including breakpoint management, variable inspection, and runtime modification that are core to our approach. The existence of established debugging protocols (e.g., Debug Adapter Protocol) facilitates cross-language integration, suggesting our methodology can extend beyond Python environments.

\paragraph{Generalizability Across Testing Frameworks.}
InspectCoder currently supports Python's unittest and pytest frameworks, which may limit applicability to projects using alternative testing paradigms. To address this, we design InspectWare with an extensible middleware architecture that abstracts debugger-specific operations and handles diverse testing scenarios. The middleware addresses potential conflicts through automatic test encapsulation, robust mock detection, and graceful handling of edge cases such as infinite loops, I/O redirection conflicts, and runtime errors. While our evaluation focuses on standard testing frameworks, the architecture is designed to be extensible to additional testing paradigms, making it adaptable to various development environments.

\subsection{Limitations and Future Work}

\subsubsection{The Debugger Incorporation Challenge: The InspectWare Solution and Beyond.}
Debuggers constitute the cornerstone of human debugging practices, enabling systematic program state inspection and root cause identification~\cite{25human-incremental-debug}. Yet their use remains sparse in modern program repair techniques, with existing approaches relying on static analysis or pre-collected traces~\cite{25AutoSD, 24NExT} rather than interactive debugging. This gap stems from two challenges: \textit{strategic action orchestration} (determining optimal breakpoints and navigation paths) and \textit{stateful session management} (where incorrect commands corrupt debugging contexts). Current agents like SWE-bench~\cite{24sweagent,autocoderover} avoid these complexities through stateless designs, sacrificing debugging's diagnostic power.

In our work, we provide an out-of-the-box solution through middleware abstraction: managing state transitions and providing strategic guidance via few-shot instruction. This enables immediate deployment without requiring specialized LLM training. While this pragmatic approach addresses immediate needs, we believe that targeted post-training—supervised fine-tuning for debugger operation and reinforcement learning for debugging efficiency—represents a promising direction for further enhancing LLMs' effective and efficient debugging capabilities. Future work will explore developing debugger-native LLMs through such specialized training, enabling them to acquire sophisticated debugging strategies (binary search localization, differential debugging, backward slicing) and native debugging intuitions that transcend middleware limitations.

\subsubsection{Dynamic Analysis in Self-Repair and Integration into Repository-Level Scenarios.}
Our evaluation targets self-repair scenarios where LLMs debug their own generated code, addressing the growing challenge of semantic errors in LLM-generated implementations~\cite{LLM-generated-bugs}. This task assumes pre-located function-level bugs, focusing InspectCoder on statement-level fault identification and patch generation that also suites the core usecases in debuggers. Unlike repository issues with ambiguous requirements, self-repair confronts \textit{specification-implementation gaps} where requirements are explicit through tests and docstrings but implementations contain logic errors. These well-defined specifications make runtime deviations directly attributable to implementation flaws, creating ideal conditions for demonstrating dynamic analysis effectiveness.

An interesting direction is to integrate debugger and dynamic analysis into repository-level repair workflows. Current repository-level repair workflows typically begin with repository understanding through iterative search and summarization, proceed to fault localization for suspicious functions, and generate massive patch candidates for reranking and selection~\cite{autocoderover,25lingmagpt}. This pipeline fundamentally relies on \textit{static debugging}, attempting to infer bugs from static semantics while ignoring dynamic runtime information, leading to inefficient trial-and-error repair.

InspectCoder offers a promising solution by injecting dynamic analysis into these repository-agent workflows. Once repository agents identify suspicious functions through exploration, our Program Inspector can plugins and examines runtime states within each function to validate intermediate behaviors and determine fault presence. For confirmed faulty functions, the Inspector provides detailed root cause analyses including precise variable values and perturbation behaviours at failure points. These dynamic insights transform patch generation from blind attempts into informed corrections guided by actual runtime behavior. We would leave this direction as exciting future work to explore.

\section{Related Work}
\paragraph{Automated Program Repair.}
Automated Program Repair (APR) aims to automatically generate patches for identified buggy code. The field has evolved through several distinct paradigms. Early template-based methods~\cite{19Template-APR1} applied predefined repair patterns for common bugs like conditional boundary errors, but were limited by manually crafted templates. Search-based techniques~\cite{12Search-APR1} addressed this by using genetic algorithms to explore patch spaces through code mutation, while constraint-based approaches~\cite{13Constraint-APR1,17Constraint-APR2} leveraged SMT solvers to generate patches satisfying logical constraints derived from test suites. The introduction of learning-based methods~\cite{19Learning-APR2,24Learning-APR1} marked a significant shift, utilizing neural networks trained on bug-fix datasets to learn repair patterns from real-world fixes rather than relying on manual templates. Recently, the advent of Large Language Models has revolutionized the field by enabling LLM-based APR approaches~\cite{LLMAPR} that leverage pre-trained code understanding capabilities, demonstrating superior generalization without requiring extensive domain-specific training.

\paragraph{Dynamic Analysis.}
Dynamic analysis is a program examiniation technique that program execution to capturing runtime behavior such as variable states and execution paths that static analysis cannot reveal \cite{dynamicanalysis, whyprogramfail}. Debugger tools serve as fundamental dynamic analysis instruments, enabling developers to strategically place breakpoints, inspect variables, and perform step-by-step execution control for in-depth program investigation strategies, like hypothesis testing and backward/forward reasoning at critical program states \cite{25human-incremental-debug}.
Recent works have begun integrating dynamic information with LLMs to enhance program understanding \cite{24LDB,25AutoSD,25exploracoder}. However, existing APR approaches primarily feed pre-collected execution traces or fixed debugging template to LLMs rather than enabling direct debugger interaction. To our knowledge, we are the first to explore empowering LLMs to autonomously operate debugger tools for dynamic analysis and patch generation.


\section{Conclusion}
We presented InspectCoder, a novel agentic program repair system that enables LLMs to perform dynamic analysis through flexible debugger interaction. Experimental results on InspectCoder show SOTA resolve rate performance while maintaining a superior time efficiency. Our work demonstrates that enabling LLMs to adaptively inspect and modify program execution through debugger collaboration significantly enhances program repair, unveiling the potential of LLM-based dynamic analysis for automated software engineering.
\bibliographystyle{ACM-Reference-Format}
\bibliography{main}

\appendix

\end{document}